\documentclass[conference]{IEEEtran}
\IEEEoverridecommandlockouts
\usepackage{flushend}
\usepackage{graphicx}
\usepackage{algorithm}
\usepackage{diagbox}
\usepackage{booktabs,subcaption,amsfonts,dcolumn}
\usepackage{amstext}
\usepackage{amsmath}
\usepackage{amsfonts,amssymb}
\usepackage{amsmath,tabu,caption}
\usepackage{amssymb}
\usepackage[dvips]{epsfig}
\usepackage{epsfig}
\usepackage[dvips]{graphicx}
\usepackage{multirow}
\usepackage{multirow}
\usepackage[utf8]{inputenc}
\usepackage[english]{babel}
\usepackage{caption}
\usepackage{subcaption}
\usepackage{float}
\usepackage{ltablex}
\usepackage{cite}
\usepackage{hyperref}
\hypersetup{colorlinks,allcolors=black}
\usepackage{xcolor}
\usepackage{physics}
\usepackage[english]{babel}
\usepackage{cite}
\usepackage{amsmath,amssymb,amsfonts}
\usepackage{algorithmic}
\usepackage{graphicx}
\usepackage{textcomp}
\usepackage{xcolor}
\def\BibTeX{{\rm B\kern-.05em{\sc i\kern-.025em b}\kern-.08em
    T\kern-.1667em\lower.7ex\hbox{E}\kern-.125emX}}
\begin{document}

\title{Quantum code division multiple access based continuous-variable quantum key distribution
\thanks{This work was supported in part by the INSPIRE Faculty Fellowship awarded by the Department of Science and Technology, Government of India (Reg. No.: IFA22-ENG 344), and the New Faculty Seed Grant (NFSG) from the Indian Institute of Technology Delhi.

This work has been accepted for presentation at the IEEE International Conference on Quantum Communications, Networking, and Computing (QCNC 2025). Copyright may be transferred without notice, after which this version may no longer be accessible.
}
}

\author{\IEEEauthorblockN{ Shahnoor Ali$^1$, and Neel Kanth Kundu$^{1,2}$ }
\IEEEauthorblockA{\textit{$^1$Centre for Applied Research in Electronics} \\ \textit{$^2$Bharti School of Telecommunication Technology and Management} \\
\textit{Indian Institute of Technology Delhi},
New Delhi, India \\
email: shahnoorali1989@gmail.com, neelkanth@iitd.ac.in}
}

\maketitle

\begin{abstract}

In this paper, we propose a quantum code division multiple access (q-CDMA) based continuous-variable quantum key distribution (CV-QKD) system. In the proposed system, the quantum states of two senders ($\text{Alice}_{1,2}$) are chaotically encoded through chaotic phase shifters and then transmitted over a quantum channel. At the receiver, the quantum states are decoded via chaos synchronization to separate the quantum states sent by the different senders and received by the two receivers ($\text{Bob}_{1,2}$) separately. We characterize the input-output relation of the quadrature between the two senders and receivers and then analyze the secret key rate (SKR) of the q-CDMA-based CV-QKD system. Our numerical results reveal that the q-CDMA approach can significantly enhance the SKR for both users when compared to the single-user case without the q-CDMA approach.

\end{abstract}


\section{Introduction}\label{Sec:1}
With the rapid development of information and communication technologies, data security and privacy are of utmost importance in the post-quantum era \cite{malina2021post,rozenman2023quantum}. The rapid advancement in quantum computing technology poses a grave threat to the security of public key cryptography systems \cite{liu2022towards}. Shor's factoring algorithm provides a polynomial time complexity algorithm for solving the prime factorization and discrete logarithm problems. Thus, the computationally secure public key cryptography algorithms like RSA and classical key distribution algorithms like Diffie-Hellman are prone to quantum attacks once the quantum computing technology matures. Therefore, quantum-resistant cryptographic algorithms are required to secure sensitive data in future communication networks. 

Although quantum computing poses a threat to conventional cryptographic algorithms, another quantum technology known as quantum key distribution (QKD) provides a solution to this threat \cite{pirandola2020advances}. QKD uses the laws of fundamental quantum mechanics to distribute symmetric keys between two users. The symmetric keys distributed using QKD protocol between the two legitimate parties (Alice and Bob) are unknown to the eavesdropper (Eve) and any eavesdropping attack can be detected with provable guarantees using the laws of quantum mechanics. Therefore, the security of QKD protocols is guaranteed by the laws of quantum mechanics, such as the quantum no-cloning theorem and Heisenberg's uncertainty principle. Continuous variable QKD (CV-QKD) is a promising solution for integrating QKD in future telecommunication networks since it can be implemented using coherent sources and homodyne/heterodyne detectors only without requiring single photon sources and detectors \cite{pirandola2020advances}.


For widespread deployment of CV-QKD in telecommunication networks, efficient multiple access schemes are required so that multiple users can distribute secret keys using the same physical channel. Multiplexing techniques for CV-QKD systems have not yet been investigated well in the literature. Prior works have proposed frequency division multiple access (FDMA), time division multiple access (TDMA), and wavelength division multiple access (WDMA) for QKD networks \cite{razavi2012multiple}. To the best of our knowledge, code division multiple access (CDMA) has not been applied to CV-QKD yet. Some of the prior works on QKD multiplexing have investigated CDMA-based multiplexing for DV-QKD protocols only \cite{sharma2020quantum}. Prior works on CDMA-based continuous variable quantum communications have considered only entanglement distribution \cite{q-cdma}, and have not characterized the secret key rates (SKR) of CV-QKD protocols. 

In this paper, we introduce a two-user quadrature code-division multiple access (q-CDMA) network employing chaotic phase shifters. We rigorously characterize the input-output quadrature relationship between the transmitters and receivers and analyze the secret key rate (SKR) of a q-CDMA-based CV-QKD system. Numerical simulations demonstrate that the proposed q-CDMA scheme enables robust secret key transmission in highly noisy channels while effectively mitigating crosstalk between users. Moreover, our findings indicate that the proposed approach significantly enhances the SKR for both users compared to the single-user scenario without q-CDMA.

The rest of the paper is organized as follows. In Sec~\ref{Sec:2}, we introduce the q-CDMA system model that is used for CV-QKD. In Sec~\ref{Sec:3}, we present the SKR analysis for the q-CDMA-based CV-QKD system. In Sec~\ref{Sec:4}, we present extensive numerical simulations to study the SKR performance of the proposed q-CDMA based CV-QKD system with experimentally realizable parameters. Finally, in the last Sec~\ref{Sec:5}, we present the conclusion and discuss the future extensions of this work. 


\section{System Model} \label{Sec:2}

\begin{figure}[h!]
\centering
\includegraphics[width=0.5\textwidth]{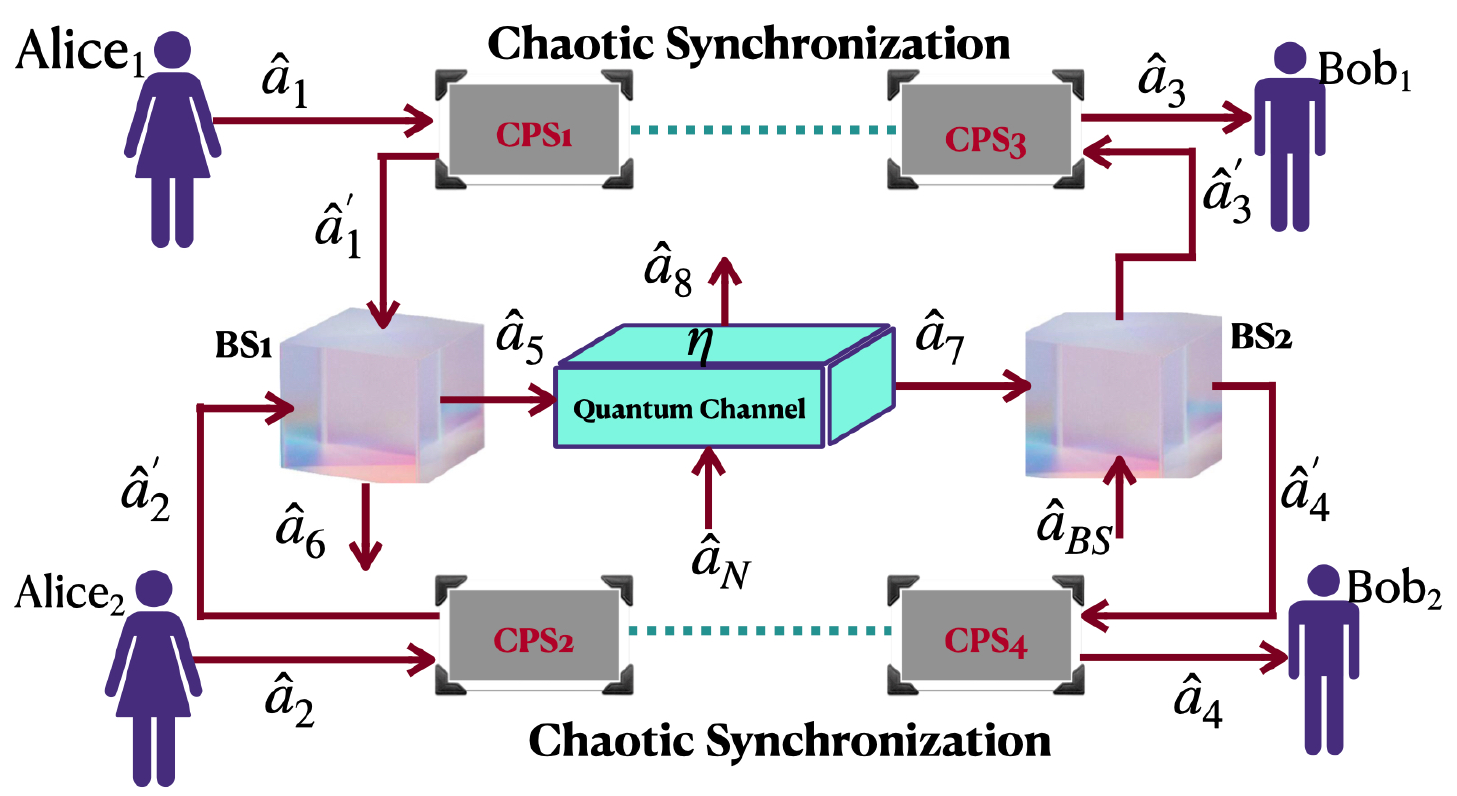}
\caption{Schematic diagram of the two users q-CDMA based CV-QKD system using chaotic phase shifters.}
\label{fig:N1}
\end{figure}

The schematic of a two-user CV-QKD network based on q-CDMA is shown in Fig.~\ref{fig:N1}. In this setup, the quantum states are initially prepared by two transmitters, $\text{Alice}_{1}$ and $\text{Alice}_{2}$, which first encode their quantum signals using two chaotic phase shifters, $\text{CPS}_{1}$ and $\text{CPS}_{2}$. The operation of these phase shifters $\text{CPS}_{1,2}$ can be modeled by the effective Hamiltoninan $\hbar\delta_{i}\hat{a}^{\dagger}_{i}\hat{a}_{i}$, where $\delta_{i}, i = 1,2$ represents the time-dependent frequency of the classical chaotic signal generated by $\text{Alice}_{1,2}$ and $\hat{a}_{1,2}$ are the annihilation operators of the signal generated by $\text{Alice}_{1,2}$. This encoding process modifies the spectral content of the quantum states over the entire spectrum. After encoding, the input states of $\text{Alice}_{1,2}$ are then multiplexed by the $50:50$ beamsplitter ($\text{BS}_{1}$), and simultaneously transmitted via a common quantum channel as shown in Fig.~\ref{fig:N1}. The eavesdropper, Eve, attempts to extract the key information by interacting with the quantum channel. At the end of the quantum channel, the quantum states are demultiplexed into two branches by a second $50:50$ beamsplitter ($\text{BS}_{2}$) and finally sent to the two receivers $\text{Bob}_{1,2}$ through two additional chaotic phase shifters $\text{CPS}_{3}$ and $\text{CPS}_{4}$. These phase shifters are used to decode the received quantum states. The operation of these chaotic phase shifters is given by Hamiltonian $-\delta_{j}\hat{a}^{\dagger}_{j}\hat{a}_{j}$ where $\hat{a}_{j}, j =3,4$ are the annihilation operators of the signal entering into $\text{CPS}_{j}$, and $\delta_{j}$ represents the frequency of the signal at $\text{CPS}_{j}$.

The action of the chaotic phase shifters $\text{CPS}_{i = 1,2,3,4}$ induce the phase shift $\exp[-i\theta_{i}(t)]$, where $\theta_{i}(t) = \int^{t}_{0}\delta_{i}(\tau)d\tau$ is the accumulated phase shift. To achieve a faithful SKR between the $\text{Alice}_{1,2}$ and $\text{Bob}_{1,2}$, the effect of $\delta_{1}$ and $\delta_{2}$ on the received quantum state must be minimized. This can be accomplished by adjusting the system parameters such that $\delta_{1} = \delta_{3}$ and $\delta_{2} = \delta_{4}$, ensuring that the phase shifts at the transmitters and receivers match \cite{q-cdma}.

To this end, we note that the proposed q-CDMA-based CV-QKD system model is inspired by the chaotic phase shift-based quantum-CDMA communication network proposed in \cite{q-cdma}. In our proposed system model, we incorporate the effect of the eavesdropper attacking the quantum channel which is specific to the CV-QKD application. Furthermore, our proposed q-CDMA-based CV-QKD system does not use linear amplification at the receiver, since it introduces additional noise, which can degrade the SKR of the CV-QKD system. 
To formalize this, we assume that the chaotic phase shifters $\text{CPS}_{1}$, $\text{CPS}_{2}$ (located at the transmitters) and  $\text{CPS}_{3}$, $\text{CPS}_{4}$  (located at the receivers) have been synchronized before initiating the key exchange between the transmitter-receiver pairs $\text{Alice}_{1,2}$ and $\text{Bob}_{1,2}$. This synchronization condition ensures that the phase shifts satisfy $\theta_{1} = \theta_{3}$ and $\theta_{2} = \theta_{4}$. Consequently, the encoding and decoding operations implemented by the chaotic phase shifters $\text{CPS}_{i=1,2,3,4}$ in Fig.~\ref{fig:N1} leads to the following input-output relations of the annihilation operators:\\
$\hat{a}'_{1} = \hat{a}_{1} e^{-i\theta_{1}}$,  $\hat{a}'_{2} = \hat{a}_{2} e^{-i\theta_{2}}$,
$\hat{a}_{3} = \hat{a}^{'}_{3} e^{i\theta_{1}}$, $\hat{a}_{4} = \hat{a}^{'}_{4} e^{i\theta_{2}}$
\begin{align}
\hat{a}_{5}&= \frac{1}{\sqrt{2}} (\hat{a}^{'}_{1} + \hat{a}^{'}_{2}),~~~\hat{a}_{6} = \frac{1}{\sqrt{2}} (\hat{a}^{'}_{1} - \hat{a}^{'}_{2})\\
\hat{a}_{7}&= \sqrt{\eta}\hat{a}_{5} + \sqrt{1-\eta} \hat{a}_{N},~~~\hat{a}_{8} = \sqrt{\eta}\hat{a}_{N} - \sqrt{1-\eta}\hat{a}_{5}
\end{align}   
where $\hat{a}_{1}$, $\hat{a}_{2}$ are the annihilation operators corresponding quantum state transmitted by $\text{Alice}_{1}$ and $\text{Alice}_{2}$, respectively. The term $\hat{a}_{N}$ represents the annihilation operator for the mode introduced by the eavesdropper, Eve, who attempts to interfere by inserting her signal into the quantum channel. The parameter $\eta$ denotes the channel transmittance, which models the losses incurred as the quantum state propagates through the channel. This transmittance can be modeled as 
$\eta = 10^{-\alpha d/10}$ where $\alpha$ is the attenuation coefficient (in dB/km), and $d$ represents the length of the optical fiber (in km) used for transmission \cite{Liu:24}. Furthermore,
\begin{align}
\hat{a}^{'}_{3}& =\frac{1}{\sqrt{2}}(\hat{a}_{7} + \hat{a}_{BS}),~~~
\hat{a}^{'}_{4} = \frac{1}{\sqrt{2}}(\hat{a}_{7} - \hat{a}_{BS})
\end{align}
denote the annihilation operators of the two output modes of the second beam splitter ($\text{BS}_{2}$) at the receiver, and $\hat{a}_{BS}$ denote the annihilation operator for the environment mode at ($\text{BS}_{2}$). The input-output relation for the two-user q-CDMA network in Fig.~\ref{fig:N1} is given by
\small
\begin{align}\label{Eq:4}
\hat{a}_{3,4}&= \frac{\sqrt{\eta}}{2}\hat{a}_{1,2} + \frac{\sqrt{\eta}}{2}\hat{a}_{2,1} e^{i(\theta_{1} - \theta_{2})} + \sqrt{\frac{1 - \eta}{2}}\hat{a}_{N} e^{i\theta_{1,2}}\\ \nonumber
&~~~~~\pm \sqrt{\frac{1}{2}}\hat{a}_{BS} e^{i\theta_{1,2}}
\end{align}
\normalsize
For a pseudo-noise chaotic phase shift $\theta_{i}(t)$, the averaging process over the broadband random signal leads to the following relation:
$\overline{\exp(\pm i\theta_{i}(t))} = \sqrt{M_{i}}$
where $M_{i}$ is the correction factor, given by
$M_{i} = \exp \left[ -\pi \int_{\omega_{li}}^{\omega_{ui}} d\omega \frac{S_{\delta_{i}}(\omega)}{\omega^{2}}\right]$
where, $S_{\delta_{i}}(\omega)$ denote the power spectral density of the signal of frequencies $\delta_{i}(t)$, where $\omega_{li}$ and $\omega_{ui}$ are the lower and upper bounds of the frequency limit of the band, respectively. The integral accounts for the spectral properties of the chaotic phase shifts, leading to the correction factor $M_{i}$ \cite{q-cdma}. This yields input-output relationship in Eq.~(\ref{Eq:4}) as
\begin{align}\label{Eq:5}
\hat{a}_{3,4}&= \frac{\sqrt{\eta}}{2}\hat{a}_{1,2} + \frac{\sqrt{M_{1}M_{2}\eta}}{2}\hat{a}_{2} + \sqrt{\frac{M_{1}(1 - \eta)}{2}}\hat{a}_{N}  \\ \nonumber
&\quad \pm\sqrt{\frac{M_{1,2}}{2}}\hat{a}_{BS} 
\end{align}
where $\hat{a}_{3},\hat{a}_{4} $ are the annihilation operators corresponding to the received modes at $\text{Bob}_{1}$ and $\text{Bob}_{2}$, respectively.

\subsection{Key Generation}
During the key generation phase, $\text{Alice}_{1,2}$ encodes the key information using Gaussian modulation. Each of them generates two statistically independent random vectors, $X_{A1,2}$ and $Y_{A1,2}$, which follow a Gaussian distribution: $\hat{Y}_{A1,2}, \hat{X}_{A1,2} \sim N(0,V_{A1,2})$, where $V_{A1,2}$ denotes the power used by $\text{Alice}_{i}$ to encode the initial key information \cite{PhysRevLett.93.170504}. Using these random variables, they prepare displaced Gaussian coherent states, denoted $\ket{\alpha_{1,2}}$, where $\alpha_{1,2} = X_{A1,2}+iY_{A1,2}$, and transmit these states through the q-CDMA network.

At the receiving end, $\text{Bob}_{1,2}$ makes quadrature measurements in the received signal modes to extract the encoded key information. The measurement can be done either by homodyne detection, which measures a single quadrature component, or by hetrodyne detection which measures both quadratures. Here, we assume that $\text{Bob}_{1,2}$ employ homodyne detection to extract real-valued measurement outcomes of one of the quadrature components. Consequently, the effective input-output relation during the key generation phase is given by the quadrature for input-output relation for Eq~(\ref{Eq:5}) 
$\{\hat{a}_{k}= (X_{k}+iY_{k})/\sqrt{2}\}_{k=1,2,3}$. For simplicity, we assume that $X_{3,4} = X_{B1,B2}$ and $X_{1,2} = X_{A1,A2}$ are the quadrature for $\text{Alice}_{1,2}$ and $\text{Bob}_{1,2}$, respectively. The quadrature measured by $\text{Bob}_{1,2}$ are given by
\small
\begin{align} 
X_{B1,2} &= \frac{\sqrt{\eta}}{2} X_{A1,2} + \sqrt{\frac{M_{1}(1 - \eta)}{2}} X_{N} + \frac{\sqrt{M_{1}M_{2}\eta}}{2} X_{I} \\ \nonumber
&~~~~~~~\pm \sqrt{\frac{M_{1,2}}{2}} X_{BS}
\end{align}
\normalsize
where $ X_{B1,2}$, represents one of the quadrature measurement outcomes obtained by $\text{Bob}_{1,2}$ while $X_{A1,2}$, represents the quadrature component transmitted of the $i$-th coherent state transmitted by $\text{Alice}_{1,2}$. Additionally, $X_{N}$ signifies the Gaussian noise quadrature introduced by an eavesdropper (Eve) into the quantum channel, and $X_{I}$ and $X_{BS}$ represent the quadrature components associated with the interference mode between $\text{Alice}_{1,2}$ and the beam splitter's environmental mode, respectively.  

The transmitted mode from $\text{Alice}_{1,2}$ has thermal noise with variance $V_{0}$, resulting in a total variance for each transmitted quadrature component given by $V(\hat{X}_{A,i}) = V_{a} = V_{S} + V_{0}$, where $V_{S}$ represents the modulation power of the initial state of  $\text{Alice}_{1,2}$. The Gaussian noise introduced by Eve has a variance of $V(X_{N})=W$, while the interference and environment quadratures have variances $V(X_{I})= \Gamma$ and $V(X_{BS})=\sigma$, respectively. Both $X_{I}$ and $X_{BS}$ follow Gaussian distributions, such that $X_{I} \sim N(0, \Gamma)$ and $\hat{X}_{BS} \sim N(0, \sigma)$. The variance $\Gamma$ of the interference quadrature can be expressed as $\Gamma = \Gamma_{0} + \Gamma_{S}$, where $\Gamma_{0}$ is the variance due to thermal noise, and $\Gamma_{S}$ is the variance arising from the interference mode.


\section{Secret Key Rate Analysis} \label{Sec:3}
In this section, we present the SKR analysis of the q-CDMA-based CV-QKD system by incorporating interference noise and environment noise. The variance of the quadrature measured by $\text{Bob}_{1,2}$, $V_{B1,2}$, \; admit
\small
\begin{align}\label{Eq:7}
V_{B1,2}&=\frac{\eta}{4}V_{A1,2} + \frac{M_{1,2}(1 - \eta)}{2}W + \frac{M_{1}M_{2}\eta}{4} \Gamma+ \frac{M_{1,2}}{2}\sigma 
\end{align}
\normalsize
where both quadrature at the transmitter side have been symmetrized; therefore, both quadrature  $X_{Ai}$, $i=1,2$ have the same variance $V_{A1}$. 


\subsection{Eve Attack Mode: Collective Attack}

The SKR analysis of CV-QKD protocols is carried out under the assumption of various potential attacks. The most general attack implemented by Eve is the collective Gaussian attack, which is modeled using the entangling cloner attack. In this attack, Eve perfectly replaced the quantum channel between $\text{Alice}_{1,2}$ and  $\text{Bob}_{1,2}$ with her own quantum channel. She then prepares ancilla modes, which are two-mode squeezed states (Einstein-Podolsky-Rosen (EPR) pair), with variance $W$. The loss of the quantum channel is simulated by a beamsplitter of transmissivity $\eta$. Eve retains one mode of the EPR pair in her quantum memory and injects the other mode into the quantum channel, resulting in the output mode $\hat{a}_{8}$ as shown in Fig~\ref{fig:N1}. Eve then collectively measures all modes stored in the quantum memory for the final coherent measurement \cite{RevModPhys.84.621}. In this paper, we use the reverse reconciliation scheme, which significantly extends the transmission distance and enables the generation of secure keys over long distances.
The secret key rate $R^{\blacktriangleleft}_{1,2}$ for the reverse reconciliation can be written as
\begin{align}
R^{\blacktriangleleft}_{1,2} :&= I(X_{B1,2}:X_{B1,2}) - \chi(X_{B1,2}:E) \label{Eq:8}
\end{align}
where $I(X_{A1,2}:X_{B1,2})$ denotes the mutual information between $\text{Alice}_{1,2}$ and $\text{Bob}_{1,2}$, and  $\chi(X_{B1,2}:E)$ represents the quantum mutual information shared between Eve and $\text{Bob}_{1,2}$.

The  mutual information based on Shannon entropy between the measurements of $\text{Alice}_{1,2}$ and $\text{Bob}_{1,2}$ denoted as $I(X_{A1,2}:X_{B1,2})$ is defined using the variances $V_{B1|A1}$ and $V_{B2|A2}$,
\begin{align}
I(X_{A1,2}:X_{B1,2})=\frac{1}{2} \log_{2}\frac{V_{B1,2}}{V_{B1,2|A1,2}} \label{Eq:17}
\end{align}
where $V_{B1,2}$ denotes the variance of $\text{Bob}_{1,2}$ given by Eq.~(\ref{Eq:7}) and $V_{B1,2|A1,2}$ represents the conditional variance given by
\small
\begin{align}
    V_{B1,2|A1,2}= \frac{\eta}{4}V_{0} + \frac{M_{1,2}(1 - \eta)}{2}W+\frac{M_{1}M_{2}\eta}{4}\Gamma_{0} + \frac{M_{1,2}}{2}\sigma
\end{align}
\normalsize

The Holevo information $\chi(X_{B1,2}:E)$ quantifies the mutual information between $\text{Bob}_{1,2}$ measurement and Eve's quantum state and is given by:
\begin{equation}
\chi(X_{B1,2}:E) := S(E)- S(E|X_{B1,2}) \label{Eq:18}
\end{equation}
where $S(E):=S(\rho_{E})$ denotes the von Neumann entropy of Eve's quantum state, and $S(E|X_{B1,2})$ represents von Neumann entropy of Eve's state conditional on $\text{Bob}_{1,2}$ homodyne measurement. Calculating these von Neumann entropies requires determining the symplectic eigenvalues $\nu_{k}$ of the corresponding covariance matrix (CMs) $\boldsymbol{\Sigma}_{E}$ and $\boldsymbol{\Sigma}_{E|X_{B1,2}}$, respectively, which describe the covariance matrix of Eve's quantum state and the conditional covariance matrix after $\text{Bob}_{1,2}$ measurement, respectively\cite{RevModPhys.84.621}.

To find the symplectic eigenvalues $\nu_{1,2}$ of $\boldsymbol{\Sigma}_{E}$, one must compute the eigenvalues of the matrix  $|i\boldsymbol{\Omega}\boldsymbol{\Sigma}_{E}|$, where $\boldsymbol{\Omega}$ is the symplectic form given by $\boldsymbol{\Omega} = \bigoplus^{2}_{k=1} \begin{bmatrix}
0 & 1 \\
-1 & 0
\end{bmatrix} $ with $\bigoplus$ denoting the direct sum of the matrices.
Eve's CM is made up from the two modes $\hat{a}^{'}_{N}$ and $\hat{a}_{N}$ which is given by
\begin{equation} \label{Eq:24}
\boldsymbol{\Sigma}_{E} = \begin{bmatrix}
E_{V} \boldsymbol{I}_2 & \Phi \boldsymbol{Z} \\
\Phi \boldsymbol{Z} & W \boldsymbol{I}_2
\end{bmatrix} 
\end{equation}
where 
\begin{equation}
    E_{V} = (1 - \eta)V_{5} + \eta W \; \; , \Phi = \sqrt{\eta (W^{2} - 1)}
\end{equation}
\begin{equation}
\boldsymbol{I}_{2} = \begin{bmatrix}
1 & 0 \\
 0 & 1
\end{bmatrix}\\, ~~\boldsymbol{Z} = \begin{bmatrix}
1 & 0 \\
 0 & -1
\end{bmatrix}\\
\end{equation}
and $ V_{5} = \frac{1}{2}(M_{1}V_{A1} + M_{2}V_{A2})$, is the variance of input mode entering into the quantum channel.
The symplectic eigenvalues of $\boldsymbol{\Sigma}_{E}$ are given by
\begin{equation}
\nu_{1,2} =  \frac{1}{2}[\sqrt{(E_{V} + W)^{2} - 4\eta(W^{2} -1)}\pm (E_{V} - W) ] \label{Eq:23}\;.
\end{equation}
Finally, the von Neumann entropy of Eve's state $S(E)$ is given by
\begin{equation}
S(E) = \sum^{2}_{k=1}g(\nu_{k}) \label{Eq:S1}
\end{equation} 
where
\begin{align}
g(x) := \Big(\frac{x+1}{2}\Big) {\rm log}_2 \Big(\frac{x+1}{2}\Big) - \Big(\frac{x-1}{2}\Big) {\rm log}_2 \Big(\frac{x-1}{2}\Big). \label{Eq:20}
\end{align}

The von Neumann entropy of Eve's state, conditioned on $\text{Bob's}_{1,2}$ homodyne measurement $S(E|X_{B})$, can be obtained by using the symplectic eigenvalues of Eve's conditional covariance matrix $\boldsymbol{\Sigma}_{E|X_{B1,2}}$. Eve's conditional covariance matrix after homodyne measurement is given by
\begin{equation}
\boldsymbol{\Sigma}_{E|X_{B1,2}} = \boldsymbol{\Sigma}_{E} - (V_{B1,2})^{-1} \boldsymbol{C} \boldsymbol{\Pi} \boldsymbol{C}^{T} 
\end{equation}
where $\boldsymbol{\Sigma}_{E}$ is defined in eq.~(\ref{Eq:24}) ,\\ $\boldsymbol{\Pi} := \begin{bmatrix}
1 & 0 \\
0 & 0
\end{bmatrix}$, 
$\boldsymbol{C}= \mqty(
\langle X_{8}X_{B1,2} \rangle \boldsymbol{I}_2 \\
\langle X^{'}_{N}X_{B1,2} \rangle \boldsymbol{Z}\\
)=\mqty(\xi_{1,2} \boldsymbol{I}_2\\
\psi \boldsymbol{Z})$ where \\ $X_{8} = \sqrt{\eta}X_{N} -\sqrt{\frac{M_{1}(1-\eta)}{2}} X_{A1} - \sqrt{\frac{M_{2}(1-\eta)}{2}} X_{I}$.

Using the above expressions, Eve's conditional covariance matrix admits
\begin{equation}
\boldsymbol{\Sigma}_{E|X_{B1,2}} = \mqty(\boldsymbol{A}&\boldsymbol{D}\\\boldsymbol{D}^{T}&\boldsymbol{B})
\end{equation}
where 
$\boldsymbol{A}= \mqty(e_{V}-\frac{\xi^{2}_{1,2}}{V_{B1,2}}&0\\0&e_{V}),~~
\boldsymbol{B} = \mqty(W-\frac{\psi^{2}}{V_{B1,2}}&0\\0&W)\\
\boldsymbol{D}= \mqty(\Phi-\frac{\xi_{1,2}\psi}{V_{B1,2}}&0\\0&-\Phi)$, and $\psi=\sqrt{(1-\eta)} \sqrt{(W^{2}-1)}$,\\ $\xi_{1,2}=\frac{1}{2}\sqrt{2M_{1,2}\eta(1-\eta)} W - \frac{1}{4}\sqrt{2M_{1,2}\eta(1-\eta)}V_{A1,2}\\~~~~~~~~~~~-\frac{1}{4}\sqrt{2M_{1,2}M^{2,1}_{2}\eta(1-\eta)}\Gamma$.

The symplectic eigenvalues of the conditional covariance matrix $\boldsymbol{\Sigma}_{E|X_{B1,2}}$ admit
 \begin{align}
\nu_{3,4}&= \sqrt{\frac{1}{2}(\Delta \pm \sqrt{\Delta^{2}-4\det \boldsymbol{\Sigma}_{E|X_{B1,2}}})}
\label{eqv3}
\end{align}
where $ \Delta = \det \boldsymbol{A} + \det \boldsymbol{B} + 2\det \boldsymbol{C}$. Therefore, the conditional von Neumann entropy of Eve's state is
 \begin{equation}
 S(E|X_{B1,2}) = \sum^{4}_{k=3}g(\nu_{k}) \label{Eq:S2}
 \end{equation}
 where $g(x) $ is define in Eq.~(\ref{Eq:20}).


The total secrete key rate (SKR) for a two users q-CDMA system based on CV-QKD is given by
\begin{align}
R^{\blacktriangleleft}&= R^{\blacktriangleleft}_{1}+ R^{\blacktriangleleft}_{2}
\end{align}
where $R^{\blacktriangleleft}_{1}$ is the SKR of $\text{Alice}_{1} -\text{Bob}_{1}$, and $R^{\blacktriangleleft}_{2}$ is the SKR of $\text{Alice}_{2}$, $\text{Bob}_{2}$ pair of users in the q-CDMA network and given by Eq~(\ref{Eq:8}).

\section{Simulation of Q-CDMA CV-QKD System}\label{Sec:4}


In this section, we numerically study the SKR performance of the two-user q-CDMA CV-QKD system. For our simulations, we assume that $\text{Alice}_{1}$ and $\text{Alice}_{2}$ utilize equal signal power to encode the initial key information. Specifically, we set $V_{A1} = V_{A2} = V_{0} + V_{S}$. As we can see from the input-output relation given by Eq.~(\ref{Eq:5}), the interference mode for the first user is $\hat{a}_{2}$ i.e., the input mode of $\text{Alice}_{2}$ and the interference mode for the second user is $\hat{a}_{1}$ i.e, the input mode of $\text{Alice}_{1}$. Therefore, we have taken the variance of the interference mode $\Gamma$ to be equal to the $V_{A1}$. The shot noise on $\text{Alice}_{1,2}$ preparation modes in CV-QKD $V_{0} = \Gamma_{0} =1$ shot noise unit (SNU) and the variance of environment noise $\sigma$ is set equal to $1$ SNU. We take the transmission coefficient of the quantum channel $\alpha=0.25$ db/Km \cite{Liu:24}. Additionally, we assumed that all $CPS_{3,4}$ are synchronized with $CPS_{1,2}$, such that the values of correction factors $M_{1}$ and $M_{2}$ are the same.

\begin{figure}[htp]
   \centering
\includegraphics[width=0.5\textwidth]{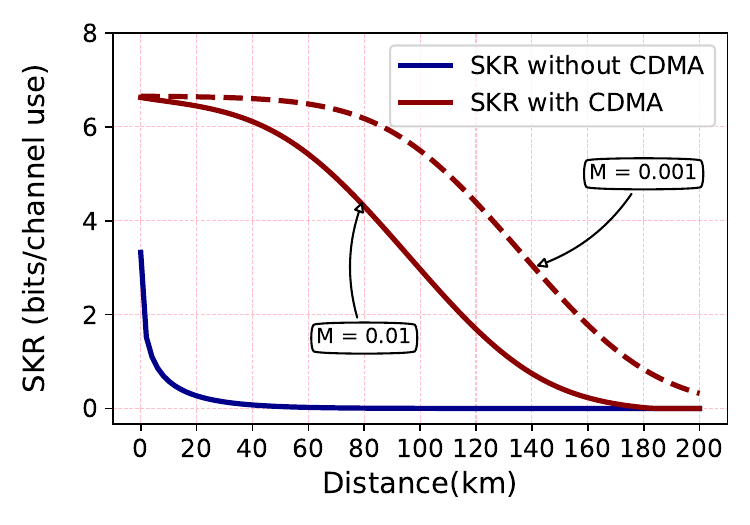}
         \caption{SKR $R^{\blacktriangleleft}$ as a function of transmission distance between $\text{Alice}_{1,2}$ and $\text{Bob}_{1,2}$, using reverse reconciliation for two different values of the correction factor $M=0.001$ (dashed line), $M= 0.01$ (solid line). Other simulation parameters are, $V_{S} =100$ SNU, $V_{0} =1.0$ SNU, $\alpha = 0.25 \; \text{dB/Km}$, and channel noise $W= \sigma = 1$ SNU.}
     \label{fig:N2}
\end{figure}

\begin{figure}[htp]
   \centering
\includegraphics[width=0.5\textwidth]{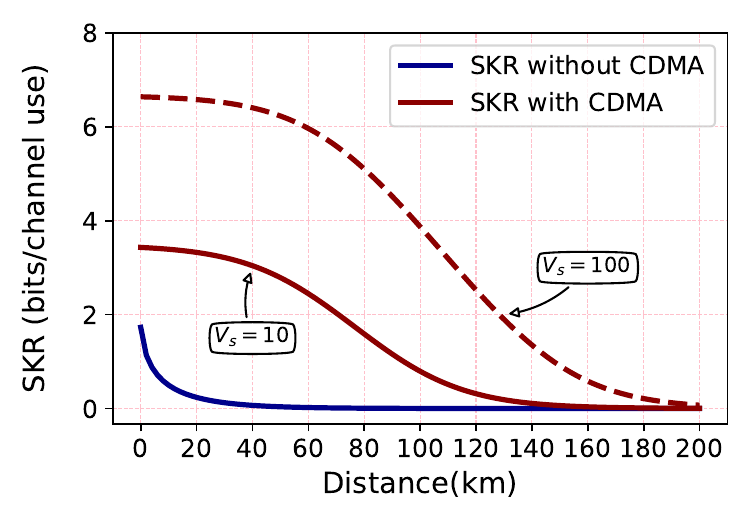}
         \caption{SKR $R^{\blacktriangleleft}$ as function of transmission distance between $\text{Alice}_{1,2}$ and $\text{Bob}_{1,2}$ for two different values of the modulation variance $V_{s}= 10$ SNU (solid line), $V_{s}= 100$ SNU (dashed line). Other simulation parameters are $V_{0} =1$ SNU, correction factor $M_{1}=M_{2} = 0.01$, $\alpha = 0.25 \; \text{dB/Km}$ and channel noise $W= \sigma = 1$ SNU.}
     \label{fig:N3}
\end{figure}

Fig.~\ref{fig:N2} shows the SKR versus transmission distance between $\text{Alice}_{1,2}$ and $\text{Bob}_{1,2}$ for the CV-QKD system with and without q-CDMA. Here, we have taken two different values of the correction factor $M_{1}=0.001$ (dashed line) and $M_{1}=0.01$ (solid line) for the q-CDMA CV-QKD system. It can be observed that the SKR decreases as distance increases for both the CV-QKD systems with and without q-CDMA. However, it can be observed that the SKR of the q-CDMA CV-QKD system is significantly higher than the baseline CV-QKD system without q-CDMA. In the absence of q-CDMA, the SKR decreases more rapidly and approaches zero. Therefore, the maximum transmission distance of the CV-QKD system can be significantly increased by using the q-CDMA approach. Furthermore, we observe that a higher value of $M_{1}$ leads to a more rapid decline in the SKR of the q-CDMA CV-QKD system. 

Fig.~\ref{fig:N3} shows the SKR versus transmission distance between $\text{Alice}_{1,2}$ and $\text{Bob}_{1,2}$ for two values of modulation variance $V_{S}=10$ SNU (solid line), $100$ SNU (dashed line). As before, it can be observed that the q-CDMA CV-QKD system has a higher SKR than the baseline CV-QKD system without q-CDMA. Furthermore, it can be observed that a higher modulation variance leads to an improvement in the SKR. Notably, without q-CDMA, the SKR diminishes more rapidly than q-CDMA.

\section{Conclusion}\label{Sec:5}
This paper presents a novel quantum code division multiple access (q-CDMA)-based multiplexing scheme for CV-QKD systems. The proposed q-CDMA-based CV-QKD system uses chaotic synchronization to encode the quantum states from multiple users and transmit quantum secure keys between two pairs of users via a common quantum channel. We model the input-output relation of the quadrature between the two senders and receivers and then analyze the SKR of the q-CDMA-based CV-QKD system by incorporating the interference and environmental noise. The numerical simulation results reveal that the q-CDMA approach can significantly enhance the SKR for both users when compared to the single-user case without the q-CDMA approach. Therefore, the proposed q-CDMA-based CV-QKD system can pave the way for the development of high-rate CV-QKD system in a multi-user setting. This paper investigated only a two-user q-CDMA CV-QKD system. Future extensions of the work should develop the system model and analyze the SKR of a general q-CDMA-based CV-QKD system with an arbitrary number of users.

\vspace{12pt}
\bibliographystyle{IEEEtran}
\bibliography{P4}

\end{document}